\documentclass[12pt,a4paper]{article}
\usepackage[left=25mm, right=25mm, top=30mm, bottom=30mm, includehead,includefoot]{geometry}
\usepackage{breakcites}

\usepackage{endnotes}

\let\footnote=\endnote

\usepackage{hyperref}

\usepackage{blindtext}
\usepackage[utf8]{inputenc}
\usepackage[brazilian]{babel}
\usepackage{amsmath}
\usepackage{amsfonts}
\usepackage{amssymb}
\usepackage[T1]{fontenc}
\usepackage{graphicx}
\usepackage{blindtext}
\usepackage{newtxtext,newtxmath}
\usepackage[nottoc]{tocbibind}

 \usepackage{setspace}
\begin{document}

\begin{center}
    
\subsection*{Proteção intelectual de obras produzidas por sistemas baseados em inteligência artificial: uma visão tecnicista sobre o tema}

Fábio Manoel França Lobato

Universidade Federal do Oeste do Pará

\{fabio.lobato@ufopa.edu.br\}
\end{center}

\begin{abstract}
    \'E ineg\'avel a pervasividade da Intelig\^encia Artificial (IA) em nossa sociedade. At\'e nas artes a IA est\'a presente. Um caso not\'orio \'e da m\'usica "Hey Ya!" do grupo OutKast, sucesso nos anos 2000, nesta \'epoca, a ind\'ustria fonogr\'afica come\c{c}ou a tomar decis\~oes baseadas em dados para tra\c{c}ar estrat\'egias a partir de previs\~oes dos h\'abitos dos ouvintes. Este \'e apenas um dos in\'umeros exemplos de aplica\c{c}\~ao de IA nas artes. O advento da aprendizagem profunda possibilitou a constru\c{c}\~ao de sistemas capazes de reconhecer o estilo art\'istico em pinturas de forma mais acurada. A gera\c{c}\~ao de conte\'udo tamb\'em \'e poss\'ivel, por exemplo, o Deepart customiza imagens a partir de dois \textit{inputs}: 1) uma imagem a ser customizada; 2) um estilo de pintura. A gera\c{c}\~ao de m\'usicas de acordo com determinados estilos a partir de sistemas baseados em IA, tamb\'em, j\'a \'e poss\'ivel. Tais possibilidades trazem \`a baila questionamentos acerca da propriedade intelectual de tais obras. Neste ensejo, a quem pertence o direito autoral de uma obra produzida a partir de um sistema baseado em Intelig\^encia Artificial? Ao criador da IA? A empresa/corpora\c{c}\~ao que subsidiou o desenvolvimento deste sistema? Ou a pr\'opria IA enquanto entidade criadora? Este ensaio visa contribuir com uma vis\~ao tecnicista sobre a discuss\~ao da aplicabilidade de direito autoral a partir de obras produzidas por IA.
    
    \textbf{Observação: } 
Este ensaio foi apresentado como trabalho final no ``Curso Avançado em Direito Autoral \& Inteligência Artificial'' e encontra-se publicado no sítio web do Instituto Observatório de Direito autoral\footnote{ \url{https://ioda.org.br/protecao-intelectual-de-obras-produzidas-por-sistemas-baseados-em-inteligencia-artificial-uma-visao-tecnicista-sobre-o-tema}}.
\end{abstract}

\begin{abstract}

The pervasiveness of Artificial Intelligence (AI) is unquestionable in our society. Even in the arts, AI is present. A notorious case is the song "Hey Ya!" of the OutKast group, successful in the 2000s. At this time, the music industry began to make decisions based on data to strategize based on predictions of listeners' habits. This case is just one of the countless examples of AI applications in the arts. The advent of deep learning made it possible to build systems capable of accurately recognizing artistic style in paintings. Content generation is also possible; for example, Deepart customizes images from two \textit{inputs}: 1) an image to be customized; 2) a style of painting. The generation of songs according to specific styles from AI-based systems is also possible. Such possibilities raise questions about the intellectual property of such works. On this occasion, who owns the copyright of a work produced from a system based on Artificial Intelligence? To the creator of the AI? The company/corporation that subsidized the development of this system? Or AI itself as a creator? This essay aims to contribute with a technicist view on the discussion of copyright applicability from works produced by AI.
\end{abstract}

\section{Introdução}

É inegável a pervasividade da Inteligência Artificial (IA) em nossa sociedade. Um exemplo interessante trata sobre o uso da IA no combate a pandemia da COVID-19 \cite{bullock2020mapping}. Neste trabalho, Bullock e colaboradores conduziram uma revisão da literatura categorizando estudos a nível molecular, clínico e social. Em todos os níveis o uso da IA foi evidenciado, incluindo atividades como: descoberta de fármacos e desenvolvimento de vacinas; planejamento hospitalar; modelagem de perfis epidemiológicos; e difusão de desinformação em redes sociais, por exemplo.  

Até nas artes a IA está presente. Um caso notório é da música  ``Hey Ya!'' do grupo OutKast, sucesso nos anos 2000 descrito por \cite{lelandtowards}. Nesta época, a indústria fonográfica começou a tomar decisões baseadas em dados para traçar estratégias a partir de previsões dos hábitos dos ouvintes. Em um breve resumo, um programa para detecção de sucesso apontou ``Hey Ya!'' como um \textit{hit} em potencial, mas o seu lançamento não teve o impacto esperado logo de imediato. Recorreram então ao comportamento do consumidor e colocaram a música próxima entre outros \textit{hits} do momento, resultado: ``Hey Ya!'' viralizou. 

Este é apenas um dos inúmeros exemplos de aplicação de IA nas artes. O advento da aprendizagem profunda possibilitou a construção de sistemas capazes de reconhecer o estilo artístico em pinturas de forma mais acurada \cite{lecoutre2017recognizing}. A geração de conteúdo também é possível, por exemplo, o Deepart customiza imagens a partir de dois \textit{inputs}: 1) uma imagem a ser customizada; 2) um estilo de pintura \cite{mao2017deepart}. A geração de músicas de acordo com determinados estilos a partir de sistemas baseados em IA, também, já é possível \cite{briot2017deep}. 

Tais possibilidades trazem à baila questionamentos acerca da propriedade intelectual de tais obras. Neste ensejo, a quem pertence o direito autoral de uma obra produzida a partir de um sistema baseado em Inteligência Artificial? Ao criador da IA? A empresa/corporação que subsidiou o desenvolvimento deste sistema? Ou a própria IA enquanto entidade criadora? 

Este ensaio visa contribuir com uma visão tecnicista sobre a discussão da aplicabilidade de direito autoral a partir de obras produzidas por IA.  O restante do manuscrito encontra-se organizado como segue. Na Seção \ref{sec:contexto} apresento um breve contexto da IA, destacando a sua evolução conceitual. Na Seção \ref{sec:iaForte} trato sobre a fronteira de IA e a sua relação com a chamada ``IA Forte''. O processo de construção de um modelo baseado em IA é discutido na Seção \ref{sec:processo}. Por fim, algumas considerações finais são feitas na Seção \ref{sec:notasfinais}.

\section{O que é a Inteligência Artificial?}
\label{sec:contexto}

A despeito da visão unicionista e ficcional do grande público, onde o robô é uma entidade de IA única, dotado de consciência, capaz de tomar decisões racionais em nível humano, a IA de fronteira aponta para outra direção. Antes de dissertar sobre, é necessário entender os motivos pelos quais esta visão está tão arraigada. Russel Stuard e Peter Norvig destacam uma declaração de Hebert Simon em 1957: 
\begin{quote}
    ``\textit{Não é meu objetivo surpreendê-los ou chocá-los, mas o modo mais simples de resumir tudo isso é dizer que agora existem no mundo máquinas que pensam, aprendem e criam. Além disso, sua capacidade de realizar essas atividades está crescendo rapidamente até o ponto — em um futuro visível — no qual a variedade de problemas com que elas poderão lidar será correspondente à variedade de problemas com os quais lida a mente humana.}'' \cite{russell2002artificial}
\end{quote}
Prêmio Turing (1975) e Prêmio Nobel (1978), Hebert Simon contribuiu notavelmente para teorias de decisão em organizações baseadas em satisfação e com racionalidade limitada. Um de seus trabalhos seminais, o livro \textit{``The sciences of the artificial''} influenciou fortemente a área chamada de \textit{``Design theory''}, que trouxe bases científicas ao projeto de sistemas artificiais - não somente baseadas em IA, mas também, \textit{``para todos os campos \textbf{(do conhecimento)} que criam artefatos para desenvolver tarefas ou cumprir objetivos e funções''}\footnote{Original: ``\textit{All fields that create designs to perform tasks or fulfill goals and functions.}''} \cite{simon1996sciences} - Tradução e grifo do autor. 

A perspectiva de que a construção de artefatos, também chamados de sistemas artificiais - e, reforço, não necessariamente baseados em IA, pode vir acompanhado da construção de conhecimento, revolucionou alguns campos, dentre os quais a própria IA, que teve um impulso quando passou a adotar com firmeza o método científico, onde as hipóteses passaram a ser submetidas a experimentos rigorosamente projetados \cite{cohen1995empirical}. 

Além da adoção de testes estatísticos e avaliação rigorosa de desempenho, o próprio projeto dos sistemas inteligentes passou a ter métodos apropriados. Destaco aqui o processo de descoberta de conhecimentos em bases de dados, por seu termo em língua inglesa \textit{Knowledge Discovery in Databases} (KDD) proposto no trabalho seminal de Fayyad, Piatesky-Shapiro e Smyth em 1997 \cite{fayyad1996data} e uma versão mais genérica bastante utilizada em Sistemas de informação que é o \textit{Design Science Research} (DSR) \cite{hevner2010design, gregor2013positioning}. Mais a frente irei dissertar sobre tais metodologias e contextualizá-las ao projeto de sistemas inteligentes. 

Mas, o que é então a Inteligência Artificial? Empresto aqui algumas definições organizadas por \cite{russell2002artificial}: 

\begin{enumerate}
    \item \textit{``[Automatização de] atividades que associamos ao pensamento humano, atividades como a tomada de decisões, a resolução de problemas, o aprendizado...''} \cite{bellman1978introduction}
    \item \textit{``A arte de criar máquinas que executam funções que exigem inteligência quando executadas por pessoas.''} \cite{kurzweil1990age}
     \item ``\textit{O estudo de como os computadores podem fazer tarefas que hoje são melhor desempenhadas pelas pessoas.''} \cite{richartificial}
    \item \textit{``O estudo das computações que tornam possível perceber, raciocinar e agir.''} \cite{patrick1992winston}
    \item \textit{``Inteligência Computacional é o estudo do projeto de agentes inteligentes.''} \cite{poole1993probabilistic}
\end{enumerate}

Organizei os conceitos em ordem cronológica para evidenciar o distanciamento do pensamento humano e inteligência (definições 1 e 2), para a execução de atividades (definição 3) e posteriormente para o projeto de agentes inteligentes (definições 4 e 5). Essa mudança adveio, dentre outros fatores, as limitações técnicas - tanto do entendimento do pensamento humano, quanto da capacidade de modelagem de tal processo. Este redirecionamento das pesquisas em IA nos leva para um questionamento importante, a ``IA forte'' realmente existe? 

\section{A ``IA forte'' existe?}
\label{sec:iaForte}

A influência do pensamento do próprio Alan Turing, considerado o pai da IA, de Hebert Simon e outros, consolidou o conceito de uma  ``Inteligência Artificial Geral'' (IAG). De fato, a IAG foi o objetivo basilar dos primórdios da IA, onde acreditava-se que a inteligência de uma máquina poderia executar com sucesso qualquer tarefa intelectual que um ser humano pode, também chamada de ``IA Forte'' ou ``IA Completa''  \cite{baum2017survey}. Também é possível definir a IAG como o campo de pesquisa que estuda máquinas capazes de executar uma ação \textbf{(ou decisão)} superior ao intelecto humano \cite{gill2016artificial}.

Para entender o conceito de ``IA Forte'', precisamos do conceito de Teste de Turing. Em sua mais recente obra, Stuart Russel afirma:
\begin{quote}
\textit{``(...) ele (Turing) propôs um teste operacional para inteligência, chamado de jogo da imitação, que mais tarde (numa forma simplificada ficaria conhecido como teste de Turing. O teste avalia o comportamento da máquina - especificamente, sua capacidade de enganar um interrogador humano e levá-lo a acreditar que ela é humana.''} \cite{russell2019human}
\end{quote}
O mais importante, a máquina estaria dotada de \textbf{consciência}, ou seja, ela precisaria estar ciente de seus próprios estados mentais e suas ações \cite{russell2002artificial}. Em \cite{russell2019human}, o autor complementa: 

\begin{quote}
\textit{``Turing esperava direcionar a discussão para a questão de saber se uma máquina poderia se comportar de determinada maneira; e, em caso positivo - se fosse capaz, por exemplo, de discursar razoavelmente sobre os sonetos de Shakespeare e seu significado -, o ceticismo sobre a IA se tornaria insustentável. (...) Na verdade, Turing escreveu 'Não poderiam as máquinas executar coisas descritas como pensamento, mas muito diferente do que um ser humano faz?'. Outro motivo para não ver o teste como uma definição de IA é que é terrível trabalhar com essa definição, e é por isso que os mais importantes pesquisadores de IA quase não se esforçam para passar no teste de Turing.''}
\end{quote}
Avaliando a fronteira em IA, percebe-se que os esforços mais próximos de uma ``consciência'' estão em um campo chamado ``\textit{Explainable Artificial Intelligence}'' (XAI) ou em conceitos correlatos como IA interpretável ou IA responsável  \cite{adadi2018peeking}. Tal campo de pesquisa surgiu a partir de pressões sociais e, consequentemente, de agências regulatórias, de que o sistema baseado em IA fornecesse não somente a decisão mais provável (predição) quanto também uma explicação - quais motivos o levaram a tomar tal decisão. 

O livro \textit{Weapons of math destruction: how big data increases inequality and threatens democracy} apresenta um bom resumo dos motivos pelos quais a XAI é tão necessária, dentre os quais, destaco a escalabilidade, o viés inerente aos dados rotulados com decisões humanas prévias (já enviesados) e o potencial impacto catastrófico de tais sistemas  \cite{o2016weapons}. O documentário \textit{Coded Bias}\footnote{https://www.netflix.com/br/title/81328723 acessado em 30 de janeiro de 2022.}, disponível na plataforma de \textit{streaming} Netflix também é um material rico de informações sobre este tema. Encerro esta seção com a ponderação feita por \cite{russell2019human}:
\begin{quote}
\textit{``O teste de Turing não tem utilidade para a IA porque é uma definição simples e altamente condicional: depende das características imensamente complicadas e basicamente desconhecidas da mente humana, que nascem tanto da biologia como da cultura.''}
\end{quote}
\section{O processo de construção de um modelo baseado em IA}
\label{sec:processo}

\begin{figure}[!h]
        \centering
        \includegraphics[width=1\textwidth]{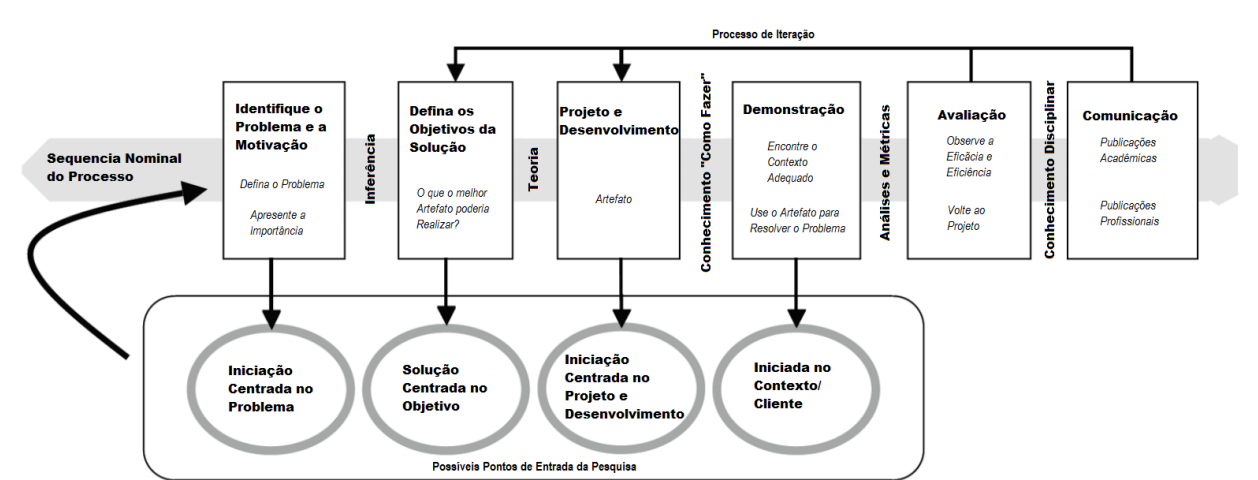}
        \caption{Etapas da metodologia DSR. Fonte: Adaptado de \cite{hevner2010design}}
        \label{fig:dsr}
    \end{figure}

Se o Teste de Turing não é o objetivo dos pesquisadores de IA, o que move este campo? Para entender, retomo o \textit{Design Science Research} e o processo de KDD, descritos por \cite{hevner2010design} e \cite{fayyad1996data} respectivamente. O DSR é um processo de seis etapas encadeadas. As fases são: i) identificação do problema e motivação; ii) definição dos objetivos da solução; iii) Projeto e desenvolvimento da solução (etapa de \textit{design}); iv) demonstração que o artefato resolve uma ou mais instâncias do problema; v) avaliação da eficiência do artefato; vi) comunicação dos achados.
Importante notar que o início do processo pode ocorrer em uma das quatro etapas iniciais, não tendo que iniciar obrigatoriamente da primeira. O desenho esquemático do DSR pode ser visto na Figura \ref{fig:dsr}. 

O interessante do DSR é que ele permite a construção de conhecimento antes, durante e após a construção do artefato, além de deixar claro os objetivos (utilidade) e desempenho (avaliação) dos constructos. Por isso esta metodologia é tão utilizada em sistemas de informação que visam a inovação tecnológica\cite{gregor2013positioning}.

\begin{figure}
        \centering
        \includegraphics[width=0.8\textwidth]{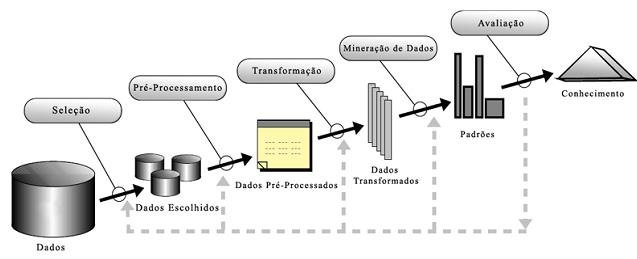}
        \caption{Etapas do KDD. Fonte: Extraído de \cite{lira2016utilizando}}
        \label{fig:kdd}
    \end{figure}

É possível notar uma clara similaridade entre o DSR e o processo de KDD, cujas etapas encontram-se descritas na Figura \ref{fig:kdd}. Enquanto o KDD é mais centrado nos dados, desde a seleção das bases até o uso do seu conhecimento - geralmente incorporado em um sistema de suporte à decisão (e.g.: análise de crédito bancário \cite{addo2018credit}; predição de suicídio baseado em dados de mídias sociais \cite{roy2020machine} \textit{etc}).

Assim como os exemplos mencionados na Introdução, o agente inteligente é projetado para um objetivo. É uma visão utilitarista e que guia o processo de avaliação do próprio modelo. Tendo como exemplo a análise de crédito bancário, a medida de desempenho adotada pode ser a taxa de perdas nos empréstimos concedidos, ou ainda, no exemplo da predição de suicídio, a taxa de acertos pelo número total de exemplos, medida chamada de acurácia. A definição das medidas de desempenho tem impacto crucial na utilidade do modelo-padrão, tal como evidenciado no supramencionado documentário \textit{Coded Bias}. 

Um processo padrão que vêm sendo aceito pela comunidade de IA é o \textit{CRoss Industry Standard Process for Data Mining} (CRISP-DM). A descrição completa do processo pode ser encontrada em \cite{wirth2000crisp}. Este processo pode ser entendido como uma evolução do KDD. A primeira etapa do CRISP-DM consiste no \textbf{entendimento do domínio de aplicação} (semelhante as duas primeiras etapas do DSR), as duas etapas subsequentes referem-se ao \textbf{entendimento} e a \textbf{preparação dos dados}. As três últimas etapas são de \textbf{construção do modelo}, o que envolve escolha de algoritmos, parametrização e treinamento; \textbf{avaliação} do modelo considerando o domínio de aplicação; e o \textbf{\textit{deployment}}, que pode ser entendido como o pleno uso do modelo. 

A literatura de IA apresenta uma vasta gama de exemplos do uso destes processos em diversos domínios de aplicação. A apresentação dos três processos teve um intuito de evidenciar o papel humano na qualidade dos modelos, sobretudo quanto ao entendimento e preparação dos dados. Estima-se que cerca de 80\% de todo o tempo do processo seja dedicado ao entendimento e preparação dos dados\footnote{https://www.forbes.com/sites/gilpress/2016/03/23/data-preparation-most-time-consuming-least-enjoyable-data-science-task-survey-says/?sh=73b421726f63 acessado em 31 de janeiro de 2022.}. 

A aprendizagem profunda ganhou notoriedade por justamente alimentar os algoritmos com os dados brutos, dispensando ou minimizando o tempo gasto na preparação dos dados, reduzindo a intervenção humana no processo - cabe ao modelo aprender a representação adequada dos dados \cite{lecun2015deep}. Destaco ainda o surgimento de uma área chamada \textit{automated machine learning}, mais conhecida pelo acrônimo \textit{automl} \cite{he2021automl}. 

De toda a sorte, os avanços na área  não apontam na direção de ausência de intervenção humana. Apesar do caso notório do AlphaZero, que por meio do paradigma de aprendizagem por reforço foi capaz de criar seus próprios dados de partidas e refinar o modelo  \cite{silver2018general}, a aplicação em mundo real de tal abordagem ainda parece incipiente - dado a natureza do ambiente dos agentes inteligentes - que não possui regras estáticas e ambiente controlado tal como um tabuleiro de xadrez ou de go. 

Em resumo, a vasta maioria das aplicações envolvendo IA ainda é humano-dependente em todo o seu processo, sobretudo na definição das medidas de desempenho e na validação dos constructos - incluindo os projetos envolvendo a criação de artefatos passíveis de proteção por direitos autorais como imagens digitais, música, poemas \textit{etc}. Portanto, quais aspectos estão relacionados à criação de uma obra por um sistema baseado em IA? Tento responder este questionamento, relacionando alguns aspectos pertinentes a criação de artefatos a partir de sistemas baseados em IA com o direito autoral, ressalto, a partir de uma visão tecnicista.

\section{Considerações finais}
\label{sec:notasfinais}

Ao longo deste ensaio busquei lançar luz ao conceito de inteligência artificial que representasse tanto o estado da arte quanto o estado da prática. Evidenciei que operacionalização de uma IA consciente e senciente ainda está restrita ao universo ficcional, e que o processo de criação de modelos-padrão - isto é, que atendam à um propósito específico (e.g.: criação de vídeos realísticos como \textit{deepfake} ou geração de imagens estilizadas como \textit{DeepArt}) dependem da mobilização de uma grande quantidade de especialistas, como analistas de banco de dados, estatísticos, programadores, analistas de desempenho \textit{etc}. Na subseção a seguir apresento os principais aspectos correlatos e levanto alguns questionamentos relacionado a proteção por direito autoral. Em seguida, traço um paralelo com outras aplicações que são tuteladas por direito autoral, encerrando com um posicionamento sobre a IA enquanto autora.

\subsection*{Aspectos relacionados à sistemas baseados em IA}
\begin{itemize}

    \item \textbf{Componente humano: } como visto, o componente humano é de extrema importância para o sucesso de um projeto de IA. É dele que vem a compreensão do domínio de aplicação que permitirá definir a(s) medida(s) de desempenho do agente inteligente, além da escolha do(s) algoritmo(s) e suas parametrizações.  Negligenciar o papel do especialista humano, tanto que a política nacional de inteligência artificial do Chile \footnote{https://minciencia.gob.cl/areas-de-trabajo/inteligencia-artificial/politica-nacional-de-inteligencia-artificial/ acessado em 21 de dezembro de 2021.}, lançada em 2021, coloca o desenvolvimento de talentos como o primeiro eixo dos fatores habilitantes de tal ação. Cabe lembrar que a IA não criaria uma obra por livre e espontânea vontade, ele é um sistema reativo, ou seja, precisa de um \textit{input} humano para que isso ocorra. Então, questiono, como podem os desenvolvedores e analistas serem deixados de lado do direito autoral advindo de uma obra produzida por IA? 
    
    \item \textbf{Plataforma de desenvolvimento (\textit{software}): } uso aqui o termo plataforma ao invés de ambiente, pois este termo é mais amplo, abarcando não somente as ferramentas utilizadas para desenvolver a aplicação como as \textit{ Integrated Development Environment} (IDE), mas também os \textit{frameworks} que encapsulam os algoritmos de IA. Exemplos de \textit{frameworks} utilizados para aprendizagem profunda incluem Keras\footnote{https://keras.io/ acessado em 31 de janeiro de 2022} e Tensorflow\footnote{https://www.tensorflow.org/ acessado em 31 de janeiro de 2022}, por exemplo. Tais plataformas podem ser protegidas por direito autoral, por meio do registro de programa de computador). Em alguns países, a proteção também pode se dar por meio de patente de software. Nota-se que algumas das ferramentas são \textit{open-source}. O que abre o questionamento, como fica a relação do licenciamento destas ferramentas e o esquema de proteção de obras construídas a partir delas?
    
    \item \textbf{Infraestrutura computacional (\textit{hardware}): } os métodos/algoritmos de IA que permitem a geração de obras passíveis de proteção por direitos autorais como músicas, desenhos e textos, requerem alto poder de processamento. Dentre tais métodos, destaco as \textit{Generative Adversarial Networks} (GANs) \cite{creswell2018generative} e a \textit{Long Short-Term Memory} (LSTM) \cite{mao2018deepj}. Apesar do barateamento do custo do \textit{hardware}, aceleração por meio de \textit{General-Purpose computing on Graphics Processing Units} (GPGPUS) e até mesmo uso de infraestrutura sob demanda como o Google Colab\footnote{https://research.google.com/colaboratory/intl/pt-BR/faq.html acessado em 31 de janeiro de 2022}, permitindo que desenvolvedores independentes consigam rodar seus códigos a baixo custo, usualmente existem empresas/companias/universidades que subsidiam times de desenvolvimento e provêm a infraestrutura necessária. Dessa forma, pondera-se que a proteção do tipo \textit{work for hire} também seja ventilada em tais casos;
    
     \item \textbf{Dados: } algoritmos baseados em aprendizagem profunda são conhecidos por requererem uma quantidade massiva de dados a fim de dispensar o processo de preparação destes. Ora, as bases de dados podem ser protegidas por direito autoral, o que levanta um questionamento, como fica o direito autoral das obras que foram baseadas em uma base de dados protegida? 
\end{itemize}

\subsection*{Paralelos com outras formas de proteção e implicações}

Alguns dos questionamentos levantados na subseção anterior já norteiam algumas possibilidades para tutelar as obras produzidas por um sistema baseado em IA.  Importante destacar que já existem artefatos criados parcialmente por sistemas baseados em IA e que são tutelados pelo direito autoral, como por exemplo, a topologia de circuitos integrados, protegido pela modalidade de propriedade intelectual \textit{Sui generis} de acordo com a Lei 11.484, de 31 de Maio de 2007. Digo que a topologia de circuitos integrados é criada parcialmente pois a IA não é capaz de criá-la sem o \textit{input}, supervisão e validação de especialistas humanos.

Outro paralelo que pode ser feito é com a produção audiovisual, considerando os inúmeros atores necessários para a construção de um sistema baseado em IA, conforme apresentado na Seção anterior. O Prof. Dr. Marcos Wachowicz apresenta no canal do Instituto Observatório do Direito Autoral no YouTube um bom contexto sobre a aplicabilidade do direito autoral à produção audiovisual\footnote{https://youtu.be/b0lWzkfVmu4 acessado em 31 de janeiro de 2022.}. 

Argumento contra a atribuição de autoria plena ao agente dotado de IA perguntando: Caberia uma penalidade a IA em caso de infração, como plágio? Isso é impensável. Desligá-la não garantiria que uma cópia do modelo fosse utilizada em outro computador. Comparo a IA a um cão, quando este ataca um ser humano - a legislação prevê que o dono do animal possa ser penalizado pela atitude do cão. Por analogia, a punição caberia aos construtores do agente inteligente (conjunto de pessoas ou corporações) - isso teria um efeito regulatório mais eficaz que o ``desligamento da máquina''.

Encerro este ensaio enaltecendo a necessidade de mais discussões sobre este tema tão atual e relevante economicamente, tanto que a comunidade científica já iniciou estudos sobre aspectos estéticos e percepções públicas acerca de criações artísticas puramente humanas e mediadas por inteligência artificial \cite{hong2019artificial}. Ressalto ainda que a regulação precisa considerar os aspectos técnicos, e ainda, que reflitam os interesses sociais e econômicos, a fim de assegurar os investimentos na área por meio da exploração da propriedade intelectual. 

\subsection*{Agradecimentos}

Este trabalho foi parcialmente financiado pelo Conselho Nacional de Desenvolvimento Científico e Tecnológico (CNPq) - DT-308334/2020-5.

\theendnotes

\bibliographystyle{apalike}
\bibliography{ioda}
\end{document}